\documentclass[aps,prb,reprint,superscriptaddress,amsmath,amssymb,preprintnumbers,floatfix,frontmatterverbose,nofootinbib,nobibnotes,bibnotes,showkeys]{revtex4-2}
\preprint{APS/000-000}
\usepackage{graphics,epsfig,graphicx,color,xcolor,epstopdf}
\usepackage[utf8]{inputenc}
\usepackage[T1]{fontenc}
\usepackage{amsmath}
\usepackage{amsfonts}
\usepackage{mathtools}
\usepackage{mathrsfs}
\usepackage{amssymb}
\usepackage{times}
\usepackage[linkcolor=blue,urlcolor=blue,citecolor=blue,colorlinks=true]{hyperref}
\usepackage{soul}
\setulcolor{red}
\definecolor{mygreen}{rgb}{0.0,0.65,0.0}
\usepackage[normalem]{ulem}
\usepackage{soul}
\usepackage{xcolor}
\definecolor{lightcyan}{RGB}{200,230,255} % celeste chiaro
\sethlcolor{lightcyan}

\newcommand{\ignore}[1]{}

\begin{document}

\title{Epitaxial growth of $\beta$--bismuthene on Sb$_2$Te$_3$}

\author{\firstname{Giorgia}~\surname{Sementilli}}
\affiliation{Istituto di Struttura della Materia-CNR (ISM-CNR), Via del Fosso del Cavaliere, 00133 Roma, Italy}
\affiliation{Startnetics-Department of Chemical Science and Techologies, University of Rome ``Tor Vergata'', Via della Ricerca Scientifica 1, 00133 Roma, Italy}

\author{\firstname{Arslan}~\surname{Masood}}
\affiliation{Istituto di Struttura della Materia-CNR (ISM-CNR), Via del Fosso del Cavaliere, 00133 Roma, Italy}
\affiliation{Dipartimento di Scienze di Base e Applicate per l'Ingegneria, Università la Sapienza Roma, Via A. Scarpa, 16 00161 Roma, Italy}

\author{\firstname{Fabio}~\surname{Ronci}}
\affiliation{Istituto di Struttura della Materia-CNR (ISM-CNR), Via del Fosso del Cavaliere, 00133 Roma, Italy}

\author{\firstname{Stefano}~\surname{Colonna}}
\affiliation{Istituto di Struttura della Materia-CNR (ISM-CNR), Via del Fosso del Cavaliere, 00133 Roma, Italy}

\author{\firstname{Marilena}~\surname{Carbone}}
\affiliation{Startnetics-Department of Chemical Science and Techologies, University of Rome ``Tor Vergata'', Via della Ricerca Scientifica 1, 00133 Roma, Italy}
\affiliation{Istituto di Struttura della Materia-CNR (ISM-CNR), Via del Fosso del Cavaliere, 00133 Roma, Italy}

\author{\firstname{Marco}~\surname{Papagno}}
\affiliation{Laboratorio di Spettroscopia Avanzata dei Materiali, STAR IR, Via Tito Flavio, Università della Calabria, 87036 Rende (CS), Italy}
\affiliation{Dipartimento di Fisica, Universit\`{a} della Calabria, Via P. Bucci, I-87036 Rende (CS), Italy}

\author{\firstname{Ziya S.}~\surname{Aliev}}
\affiliation{Baku State University, AZ1148 Baku, Azerbaijan}
\affiliation{Institute of Physics, Ministry of Science and Education of Azerbaijan, AZ1143 Baku, Azerbaijan}

\author{\firstname{Evgueni V.}~\surname{Chulkov}}
\affiliation{Departamento de Polímeros y Materiales Avanzados: Física, Química y Tecnología, Facultad de Ciencias Químicas, Universidad del País Vasco UPV/EHU, 20080 San Sebastián/Donostia, Spain}
\affiliation{Donostia International Physics Center (DIPC), 20018 Donostia-San Sebastián, Basque Country, Spain}
\affiliation{Centro de Física de Materiales (CFM-MPC), Centro Mixto CSIC-UPV/EHU, 20018 Donostia-San Sebastián, Basque Country, Spain}
\affiliation{Laboratory of Electronic and Spin Structure of Nanosystems, Saint Petersburg State University, 198504 Saint Petersburg, Russia}

\author{\firstname{Sergey V.}~\surname{Eremeev}}
\affiliation{Institute of Strength Physics and Materials Science, Siberian Branch, Russian Academy of Sciences, 634055 Tomsk, Russia}
\affiliation{Laboratory of Electronic and Spin Structure of Nanosystems, Saint Petersburg State University, 198504 Saint Petersburg, Russia}

\author{\firstname{Roberto}~\surname{Flammini}}
\email{roberto.flammini@cnr.it}
\affiliation{Istituto di Struttura della Materia-CNR (ISM-CNR), Via del Fosso del Cavaliere, 00133 Roma, Italy}

\date{\today}% It is always \today, today,

\begin{abstract}
    Over the past decades, two-dimensional crystals have attracted considerable interest as promising materials for electronic and optoelectronic applications. Among them, graphene analogs composed of heavy atoms occupy a particularly distinctive niche due to their enhanced spin–orbit interaction. Here, we present an epitaxial heterointerface formed by $\beta$-bismuthene on Sb$_2$Te$_3$, a well--known three--dimensional topological insulator. Using scanning tunneling microscopy, we systematically investigated the effects of Bi coverage and substrate temperature on nucleation processes, island morphology, and atomic structure. In addition, substrate-induced defects were identified throughout the bismuthene lattice.
\end{abstract}

\maketitle

\section{Introduction}

Bulk bismuth is a well-known topological semimetal \cite{HasanAPS2010,Schindler2018,KoroteevPRL2004,LvRevModPhys2021}, preserving the non--trivial electronic structure even in its 2-dimensional allotropes \cite{akturk_single_2016}. As a matter of fact, both $\alpha$-- and $\beta$--bismuthene were predicted \cite{akturk_single_2016,HoganPRB2021} and subsequently grown on several substrates \cite{lu_NL_2015, LYU2021, gou_2dimensional_2023, ReisS2016}.
In particular, a Bi bilayer (BLs) with a buckled honeycomb structure, commonly referred to as $\beta$-bismuthene, garnered a considerable interest being proposed as 2D quantum spin Hall material \cite{MurakamiPRL2006, WadaPRB2011}. 

The first experimental realization of $\beta$--bismuthene was reported by Reis et al., who successfully obtained a single Bi BL on SiC(0001) \cite{ReisS2016}. This result represented a major milestone, demonstrating for the first time that a two--dimensional (2D), topologically nontrivial phase of Bi could be experimentally stabilized with sufficient crystalline quality to enable detailed investigations of its electronic, structural, and topological properties.
Subsequently, it was shown that $\beta$--bismuthene could also be realized on various conventional substrates, such as Si, HOPG or graphene \cite{JiangPRM2022,KowalczykSS2011,SongAMI2015,TilgnerNatComm2025}; however, in these systems, the appearance of the characteristic buckled honeycomb phase generally occurs only upon reaching specific critical thicknesses \cite{HOFMANNPSS2006, NagaoAPS2004, HiraharaAPS2006}. These findings clearly highlight the crucial role played by the substrate in the stabilization of $\beta$--bismuthene. 

From an experimental perspective, the ideal substrate must satisfy two apparently competing requirements: it must enable an ordered and well-controlled epitaxial growth of the adlayer, while simultaneously providing sufficient electronic decoupling from it to preserve the intrinsic properties of the 2D system \cite{Bae2019}. Within this framework, van der Waals (vdW) epitaxy emerges as a particularly effective strategy, as the weak interaction between film and substrate strongly suppresses strain effects and reduces electronic hybridization that typically characterizes conventional heteroepitaxy dominated by lattice mismatch \cite{WalshAMT2017}. A decisive step forward in this direction was achieved with the growth of Bi on three--dimensional (3D) topological insulators (TI) such as Bi$_2$Te$_3$ \cite{HiraharaPRL2011,ChenAPL2012,HiraharaPRL2012,YangPRL2012,miao_evolution_2015},   Bi$_2$Se$_3$ \cite{EichAPS2014,ZhangAPL2015,SuCM2017} and Bi$_2$Se$_2$Te \cite{KimPRB2014} as well as on the magnetic TIs of the MnBi$_2$Te$_4$ family \cite{Klimovskikh2024}. In these systems, the interface is naturally governed by vdW interactions, allowing the stabilization of Bi films already at sub-BL thicknesses and even at room temperature (RT) \cite{EichAPS2014, YangPRL2012, GaoCP2013, HiraharaPRL2012, ChenAPL2012}. 

Despite the significant progress achieved so far, many material combinations remain unexplored. In particular, to the best of our knowledge, no studies have yet reported the growth of $\beta$--bismuthene on Sb$_2$Te$_3$, a 3D TI structurally similar to Bi$_2$Te$_3$ and Bi$_2$Se$_3$ \cite{ZhangNP2009}, despite the almost perfect match between the unit cell parameters of the Sb$_2$Te$_3$ substrate and $\beta$--bismuthene adlayer. This interface could represent an interesting candidate for the realization of 2D/3D TI heterostructures with potential of optimal structural and electronic quality, possibly leading to the exploitation of proximity effects \cite{HutasoitPRB2011,MenshchikovaNL2013,EssertNJP2014,JinPRB2016, HoltgreweSR2020}.

In this work, we report the first experimental realization of $\beta$--bismuthene on Sb$_2$Te$_3$, accompanied by a systematic study of the growth mechanisms and atomic structure of the interface. Using scanning tunnelling microscopy (STM), we identified the optimal conditions for stabilizing an atomically ordered $\beta$--bismuthene/Sb$_2$Te$_3$ heterointerface.

\begin{figure}
    \centering
    \includegraphics[width=8.5cm]{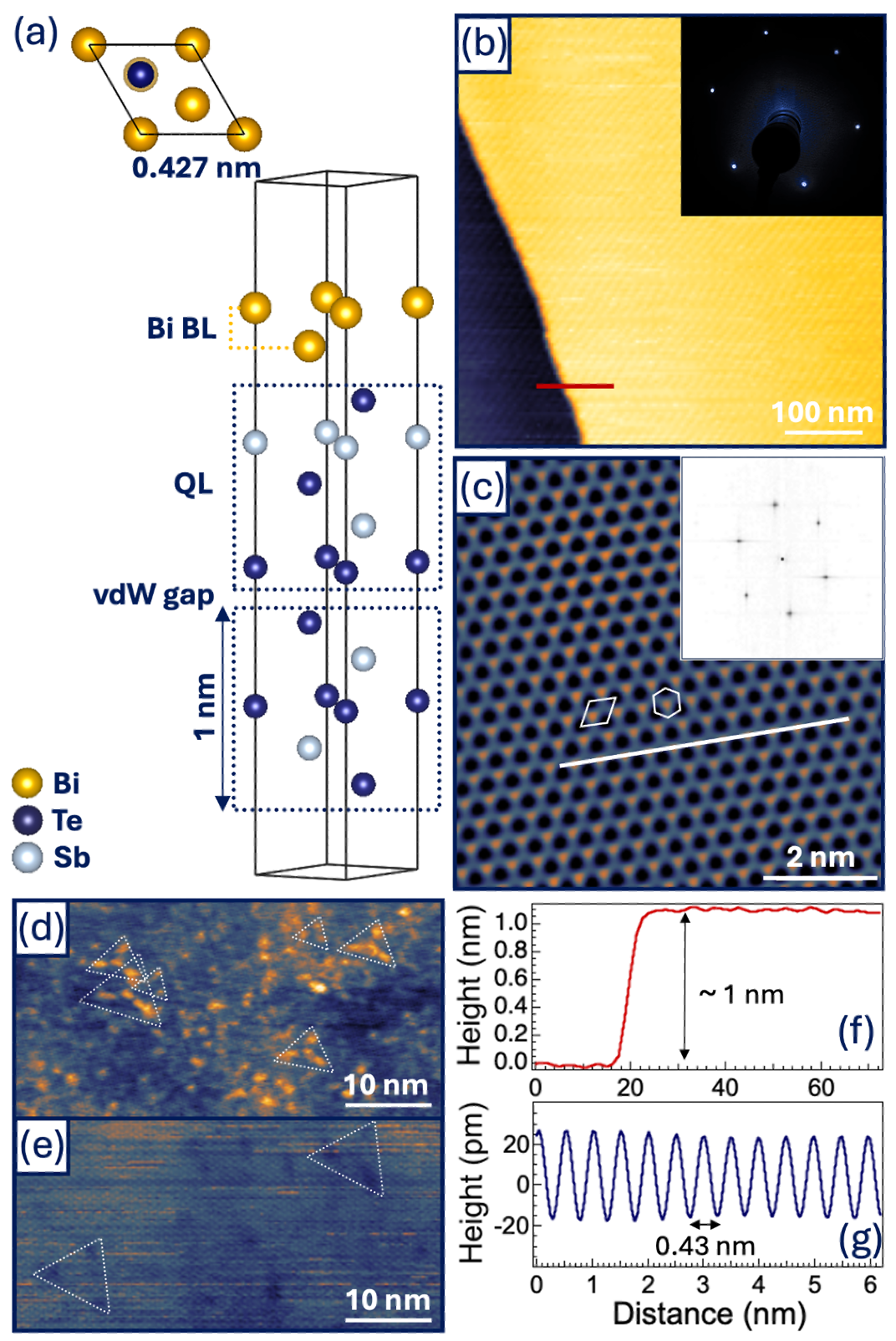}
    \caption{(a) $\beta$--bismuthene/Sb$_2$Te$_3$ structure: the substrate unit cell with the QL is highlighted. (b) STM image of the freshly cleaved Sb$_2$Te$_3$ surface, acquired with V$_b$ = +2.0 V and I$_T$ = 2.0 nA, showing a large terrace with a well-defined step edge. In the inset a LEED pattern of the surface is displayed. (c) Atomically resolved STM image of the Sb$_2$Te$_3$ surface (V$_b$ = +0.035 V and I$_T$ = 5 nA), processed using a 2D FFT filter shown in the inset (10$\times$10 nm$^{-2}$) \cite{SupplementalMaterial}. (d,e) STM images (50$\times$25 nm$^2$, taken not in the same area) of the Sb$_2$Te$_3$ surface highlighting point-like defects. In (d), the defects appear as bright protrusions (V$_b$ = +0.7 V, I$_T$ = 2.2 nA), whereas in (e) they show up as dark depressions (V$_b$ = +0.4 V, I$_T$ = 1.8 nA). (f) Apparent height profile extracted along the line indicated in panel (b); (g) Profile extracted from panel (c), along a row of atoms. All STM images were acquired at 80 K.}
    \label{1}
\end{figure}

\section{Materials and methods}

Bulk single crystals of Sb$_2$Te$_3$, oriented along the (0001) direction were grown using the vertical Bridgman method and characterized by X-Ray Diffraction (XRD) \cite{AlievPRB2015}. The samples were exfoliated in ultra-high vacuum (UHV) conditions from the bulk crystal, with lateral dimensions of 4$\times$3 mm$^2$ and a thickness of approximately 2 mm. The experiments were performed in a UHV system consisting of two interconnected chambers (working pressure 10$^{-10}$ mbar): a preparation chamber equipped with low-energy electron diffraction (LEED) and Auger electron spectroscopy (AES), and an analysis chamber equipped with an Omicron LT-STM scanning tunneling microscope.

Bi was deposited onto the Sb$_2$Te$_3$ surface by evaporation from a temperature-controlled Knudsen cell. Depositions were carried out at different coverages and substrate temperatures. At RT and 200 K a deposition rate of 0.06 \AA/min was used, while a rate of 0.7 \AA/min was employed at 150 K. All deposition rates were calibrated beforehand using a quartz crystal microbalance (QCM). The overall uncertainty in the deposition rate, accounting for the QCM accuracy, geometric factors, and the stability of the Knudsen cell, is on the order of 10\%. In the following, the term bismuthene will be used to refer exclusively to $\beta$-bismuthene. The nominal coverage was determined by considering a bismuth atom density of $6.54 \times 10^{14}$ atoms/cm$^2$ that is, 2 Bi atoms per antimony telluride surface unit cell. In this way, the nominal coverage is reported directly in BL, allowing an immediate and consistent comparison between different depositions.

STM images were acquired with the sample maintained at 80 K using liquid-nitrogen cooling bath. STM tips were mechanically cut from a PtIr wire, and the scanner was calibrated using the clean Sb$_2$Te$_3$ surface. The applied bias voltage is referenced to the sample, with positive (negative) bias corresponding to empty (occupied) electronic states. Images were acquired at both positive and negative bias; since no significant differences were observed, only images acquired at positive bias are shown for clarity. STM images were processed using Gwyddion\textsuperscript{\textregistered} software \cite{NecasJP2011}.

\section{Results and discussion}

%substrate

Clean and atomically flat Sb$_2$Te$_3$ surfaces were obtained by \emph{in--situ} exfoliation, exploiting the naturally layered structure of the material \cite{ZhangNP2009}, which consists of weakly coupled quintuple layers (QLs) (Fig. \ref{1} panel (a)). Extended terraces, separated by steps of approximately 1 nm apparent height were obtained (Fig. \ref{1} panels (b) and (f)), corresponding to the thickness of a single QL \cite{abrikosov1976production,Anderson:a11041,WangNR2010}. The resulting surface is Te-terminated and retains an unreconstructed (1$\times$1) structure \cite{UrazhdinPRB2002,ZhangPRL2009}, with atoms arranged in a hexagonal lattice and an interatomic distance of approximately 0.43 nm (Fig. \ref{1} panel (g)) \cite{JiangPRL2012, SinhaSI2025}.

The elemental composition of the surface was verified by AES \cite{SupplementalMaterial}. Within the sensitivity of the technique, no traces of oxygen or surface oxides were detected, indicating the absence of oxidation after cleavage. A weak carbon signal was, however, observed, consistent with previous reports and commonly attributed to intrinsic residues from the growth process \cite{PlucinskiJAP2013}.
The LEED pattern shows a sharp and well-defined hexagonal arrangement of diffraction spots, consistent with the symmetry of the Sb$_2$Te$_3$ surface (inset Fig. \ref{1} (b)). 

%Defects

In panels (d) and (e) of Fig. \ref{1} STM images of the Sb$_2$Te$_3$ surface are displayed, in which the presence of native point defects is clearly visible. Such native defects are commonly found in 3D TI based on Te and Se, and their presence is attributed to intrinsic factors related to the crystal growth conditions \cite{NetsouACSN2020,DaiAPS2016}. Defects locally perturb the local density of electronic states (LDOS) and appear in STM images as triangular features, observable either as protrusions or depressions, depending on the nature of the defect and the applied bias.

The triangular appearance of the defects is strongly influenced by the chemical bonding within the QLs. In contrast to the weak interaction between adjacent QLs, the bonding within each QL (with stacking sequence Te--Sb--Te--Sb--Te) is strong and dominated by the $p$ orbitals of Sb and Te atoms. These orbitals are intrinsically directional, and their overlap is therefore highly anisotropic \cite{LiuAPS2010}. In particular, the $p$ orbitals overlap preferentially along specific crystallographic directions, giving rise to what are commonly referred to as $pp$ chains which, guided by the  symmetry of the surface lattice, result in the triangular defects observed in the STM images. As reported by Jiang et al. \cite{JiangPRL2012}, Sb$_2$Te$_3$ can exhibit a total of five main types of intrinsic defects, including antisites, vacancies, and more complex configurations, whose relative populations depend on the crystal growth conditions and the formation energies of each defect.

In our work, two main types of native defects were identified: Sb antisites and Te vacancies. Their identification is facilitated by the distinct contrast clearly visible in the STM images (Fig. \ref{1}) . Sb antisites, that is, Sb atoms substituting for Te atoms in the lattice, appear as triangular protrusions (Fig. \ref{1} (d)), whereas Te vacancies manifest as triangular depressions (Fig. \ref{1} (e)). The relative visibility of the two defect types strongly depends on the applied bias: at higher biases (V$_b$ =+0.7 V), Sb antisites are more prominent, while at lower biases (V$_b$ =+0.4 V), Te vacancies become predominant \cite{JiangPRL2012}. The present work is not aimed at a systematic investigation of intrinsic defects and the analysis is restricted to the defect species most clearly resolved in our STM images, to evaluate the quality of the substrate surface. The absence of other defect types suggests that, under our experimental conditions, Sb antisites and Te vacancies constitute the dominant contributions to the surface defect density.

\begin{figure}
    \centering
    \includegraphics[width=8.6cm]{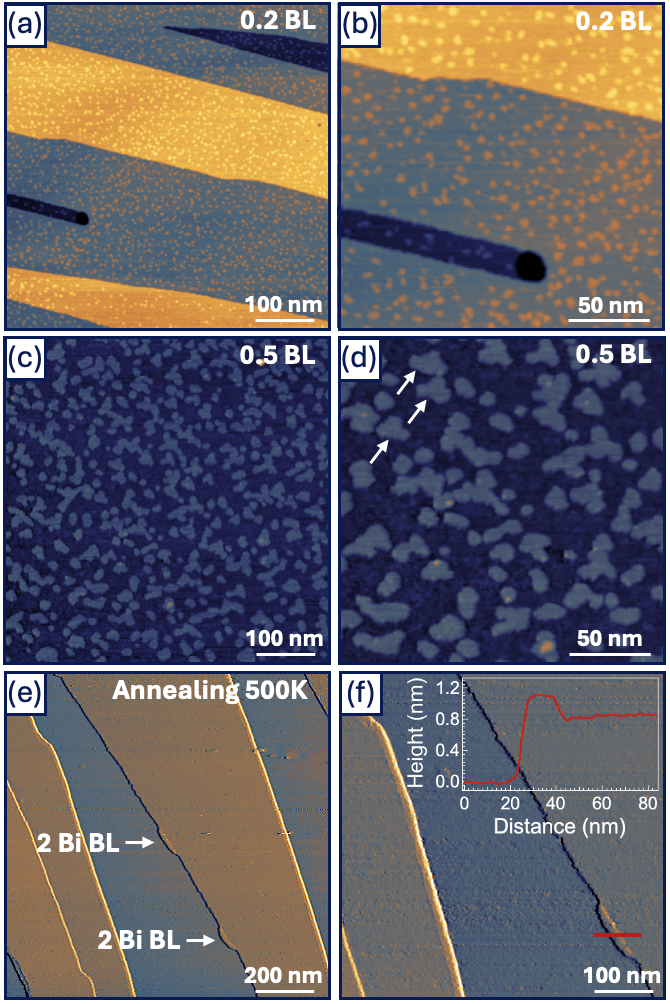}
    \caption{(a–d) STM images of Bi growth on Sb$_2$Te$_3$ at RT for coverages of 0.2 and 0.5 BL (scan areas: 500$\times$500 nm$^2$ and 200$\times$200 nm$^2$). (a,b) 0.2 BL coverage (V$_b$ = + 1.5 V, I$_T$ = 3.0 nA). (c,d) 0.5 BL coverage (V$_b$ = + 0.5 V, I$_T$ = 3.0 nA). (e) 1000$\times$1000 nm$^2$ image of the Sb$_2$Te$_3$ surface with 0.5 BL Bi coverage, after annealing at 500 K for 10 min (V$_b$ = + 0.5 V, I$_T$ = 3.0 nA). (f) Magnification of image (e) to highlight Bi accumulation. In panels (e) and (f), the images were processed by applying a derivative along the horizontal axis. In the inset of panel (f), the line profile extracted from the image before applying the derivative process is shown. All images were acquired at 80 K.}
    \label{2}
\end{figure}

To study adsorbate nucleation and growth, we monitored the evolution of the interface as a function of the amount of Bi deposited at room temperature (RT). A series of STM images for 0.2 and 0.5 BLs coverage are displayed in Fig.\ref{2}. At low coverage, the Bi islands appear as small bright protrusions on the surface, indicating the earliest stages of nucleation (Fig. \ref{2} (a-b))~\cite{JnawaliPRB2009}. At the beginning, the surface diffusion determines the competition between the nucleation of new islands and the growth of already existing ones. Up to a coverage of approximately 0.4 BL (not shown), islands density increases, nucleation being still the relevant process. 
At coverages around 0.5 BL (Fig. \ref{2} (c-d)), a clear change in the growth regime is observed: the islands begin to interact and coalesce, leading to a pronounced morphological transition. Most of the compact islands transform into irregular structures, while some start developing the characteristic triangular shape \cite{DongNC2020}. This behavior indicates that in this regime the growth is most likely driven by adatom diffusion. 

To evaluate the possibility of improving islands ordering at 0.5 BL, an annealing  at 500 K was performed for 10 min. It was observed that bismuth migrates towards the edges of the QL terraces, see panel (e) of Fig.\ref{2}, preventing the formation of regular and uniform islands. During annealing, Bi atoms acquire sufficient mobility to diffuse across the surface, but they tend to stabilize along the QL edges, where atomic coordination and local energy are more favorable \cite{ChenAPL2012}. The edges therefore act as “traps”, guiding the migration of adatoms along paths of minimum energy, possibly leading to double BL high structures, inset of panel (f). Interesting enough, we notice that no substrate disruption or formation of ulterior structures showing different apparent heights (possibly ascribable to Te or Sb segregation) was detected. 

\begin{figure*}[t]
    \centering
    \includegraphics[width=17.9cm]{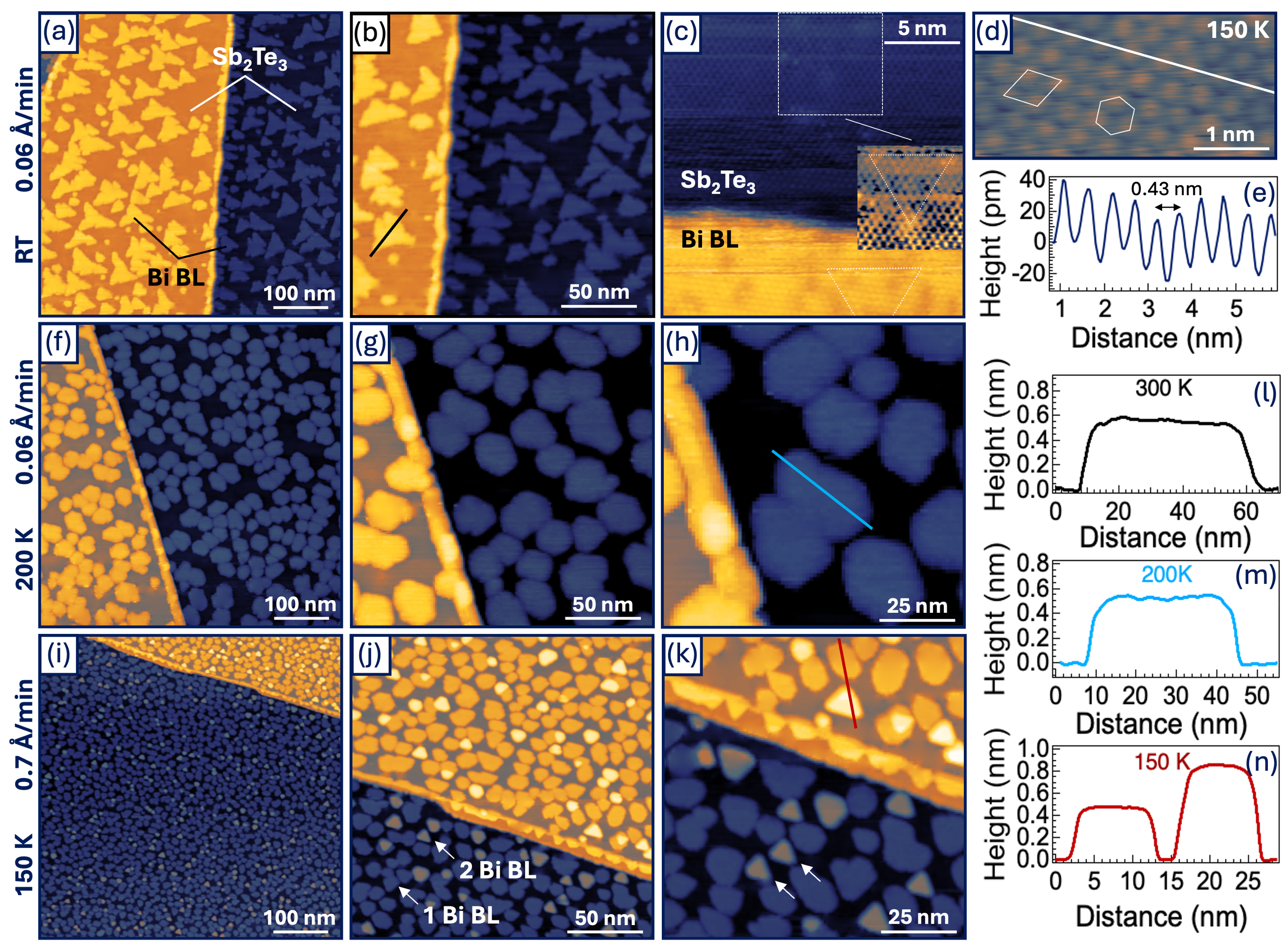}
    \caption{STM images acquired at 80 K of the Sb$_2$Te$_3$ surface upon nominal deposition of 1 BL of Bi. Sb$_2$Te$_3$ substrate appears as the darker regions, while Bi islands appear brighter. Yellowish and blueish areas highlight top and bottom of a substrate step. Panels (a) and (b) show the surface morphology for 1 BL of Bi deposited at RT, with scan areas of 500 $\times$ 500 nm$^2$, 200$\times$200 nm$^2$ (V$_b$ = +0.6 V, I$_T$ = 2.0 nA), respectively. In panel (c) image of the boundary between bismuthene and Sb$_2$Te$_3$ surface is displayed, with atomic resolution (20$\times$20, nm$^2$ V$_b$ = +0.5 V, I$_T$ = 2.0 nA). The area highlighted by the dashed square at the top is shown, at higher contrast, alongside it. The dashed triangle on top indicates a substrate defect, while the one at the bottom the effect a substrate defect on the adlayer (see text). Panel (d) shows atomic resolution within a 1 BL island (4.0$\times$1.9 nm$^2$, V$_b$=+0.05 V, I$_T$=2.3 nA), while in panel (e) it is displayed the profile of a row of atoms as extracted along the line of panel (d). Panels (f), (g) and (h) show the surface morphology for 1 BL of Bi deposited at 200 K, within scanned areas of 500$\times$500 nm$^2$, 200$\times$200 nm$^2$ and 100$\times$100 nm$^2$ (V$_b$ = +0.1 V, I$_T$ = 3.0 nA), respectively. Panels (i), (j), and (k) show the surface morphology for 1 BL of Bi deposited at 150 K, within scanned areas of 500$\times$500 nm$^2$, 200$\times$200 nm$^2$ and 100$\times$100 nm$^2$ (V$_b$ = +0.1 V, I$_T$ = 3.0 nA), respectively. Panels (l--n) show representative apparent heights of Bi islands, in the three deposition cases. Profiles color correspond to the lines color.}
    \label{3}
\end{figure*}

At a coverage of 1 BL (Fig. \ref{3}(a-b)), the islands exhibit the well-defined triangular shapes (possibly due to the formation of \textit{zig-zag} edges), with a uniform distribution \cite{GaoCP2013, YangPRL2012, ChenAPL2012}: the triangular morphology can be explained by the epitaxial relationship with the Sb$_2$Te$_3$ substrate. During growth, the islands tend to minimize edge energy: on a hexagonal lattice, triangular shapes with edges along high-symmetry crystallographic directions are more stable than irregular or rounded structures.

For 1 BL, the effect of substrate temperature on the morphology of Bi islands was further investigated, comparing depositions at RT, 200 K, and 150 K. At 200 K (Fig.\ref{3}, panels (f), (g) and (h)), unlike the bismuthene/Bi$_2$Te$_3$ and bismuthene/Bi$_2$Se$_3$ systems \cite{GaoCP2013, YangPRL2012}, the islands lose their characteristic triangular shape and their edges appear less defined: the reduced adatom mobility limits their ability to migrate and aggregate along preferential crystallographic directions, preventing the ordering necessary for the formation of well-defined triangular contours.
At 150 K, the higher growth rate brings the deposition closer to the kinetic limit and further reduces adatom mobility, resulting in a high density of small islands (Fig.\ref{3}, panels (i), (j) and (k)). Under these conditions, both 1 BL and 2 BL islands are observed. 1 BL islands exhibit irregular morphologies similar to those grown at 200 K, probably due to insufficient adatom mobility to self-organize along crystallographic directions. In contrast, the 2 BL islands show a triangular shape: the second layer grows atop pre-existing ordered islands, possibly using the defined edges of the first BL as a template. Even with limited mobility, adatoms in the second layer can aggregate along these pre--established edges, possibly stabilizing the triangular geometry. The observed morphology thus reflects the balance between nucleation and growth, governed by the larger adatom mobility on top of the 1 BL than on the substrate.

Despite the morphological difference of islands grown at different temperatures, the formation of bismuthene was confirmed in \textit{all} cases. Atomic-resolution images reveal indeed a hexagonal structure: in panel (d), the surface unit cell and a profile showing a period of nearly 0.43 nm are displayed, in perfect registry with the Te layer underneath.
This image is representative of the structure of the bismuthene islands at all temperatures. Islands apparent height of approximately 0.5 nm were observed, at the three temperatures, as indicated by the extracted profiles of panels (l--n). The apparent height is compatible with a bismuthene BL, considering the interlayer distance from the substrate, being the height of a free standing BL estimated to be around 0.39 nm \cite{RODRIGUEZMS2021, MonigPRB2005, VladislavSS2010}. The experimental value measured in our work is very close to those reported in the literature \cite{EichAPS2014, ZhangAPL2015}. These results highlight how the substrate temperature affects the size, shape, and distribution of the islands, while still allowing the formation of bismuthene.

As further evidence of the epitaxial growth of Bi on Sb$_2$Te$_3$, an interfacial region between bismuthene and the substrate has been imaged with atomic resolution (Fig.\ref{3} (c)). The upper part of the image corresponds to the substrate, while the lower part shows the bismuthene adlayer. Both regions exhibit atomic resolution; however, the boundary is not definite due to the enhanced local mobility of atoms in the vicinity of the step edge. In these boundary regions, atoms have a lower coordination number than atoms on extended terraces. They are more weakly bound and therefore more susceptible to surface diffusion and STM-tip-induced rearrangements. As a consequence, the interface can appear blurred on a timescale comparable to the STM acquisition time, preventing a sharp topographic definition of the boundary. Nevertheless, a perfect lattice matching between the two regions is clearly observed, indicating a coherent epitaxial relationship between adlayer and substrate.

Furthermore, in the lower part of Fig. \ref{3}(c) (on bismuthene), one dashed triangle highlights a defect with size and morphology compatible with those present on the pristine substrate surface (see dashed triangle on the top part of the image). This correspondence suggests that local perturbations of the electronic density of states induced by substrate defects may propagate into the adlayer, thereby influencing its local electronic structure and leading to a replication of the defect signatures within the bismuthene layer, as reported also for $\beta$--antimonene BLs \cite{FlamminiNT2018, HoganACSNano019}. The nature and type of substrate defects may therefore represent an additional tuning knob for the deliberate tailoring of the potentially topologically non-trivial electronic structure of the interface.

The realization of an atomically sharp and fully epitaxial interface between $\beta$-bismuthene and Sb$_2$Te$_3$, as demonstrated in this work, can provide a unique platform for the study of low-dimensional electronic states. In particular, the high structural quality and the excellent lattice matching are key elements that can facilitate the interpretation of complex electronic structures \cite{FlamminiAPR2025}.

\section{Conclusions}

The epitaxial growth of $\beta$--bismuthene on Sb$_2$Te$_3$ interface was experimentally realized. From the study of the initial nucleation stages, we identified a minimum coverage of approximately 0.5 BL necessary to trigger island coalescence and the formation of triangular islands, onset of $\beta$--bismuthene formation. Furthermore, at a nominal coverage of 1 BL, we observed that varying the substrate temperature during deposition from RT to 150 K changes the morphology, size, and density of the islands, while still preserving the hexagonal honeycomb structure. Overall, these results establish that $\beta$--bismuthene/Sb$_2$Te$_3$ can be considered a prototypical van der Waals heterostructure, ideal for studying and controlling the interaction between two non--trivial electronic structures.  

\acknowledgments
The technical support by M. Rinaldi, G. De Santis and G. Emma is warmly acknowledged. E.V.C. and S.V.E. acknowledge Saint-Petersburg State University for the research project 125022702939-2. S.V.E. acknowledges support from the Government research assignment for ISPMS SB RAS, Project No. FWRW-2026-0008. 
\section*{References}
\bibliographystyle{apsrev4-2}
\bibliography{biblio_Bi}

\end{document}